\pdfoutput=1
\documentclass[a4paper]{article}
\usepackage{authblk}
\usepackage[margin=1in]{geometry}

\date{}
\usepackage{graphicx,hyperref}

\usepackage{amsmath,amsfonts,amssymb}

\usepackage{physics}
\usepackage{comment}

\newcommand{\LieD}{\mathcal{L}}

\newcommand{\vardiff}[2]{\tilde{\delta}_{#1}{#2}}

\newcommand{\bb}{b}
\newcommand{\mm}{m}

\newcommand{\dphi}[1]{\var{\phi_{(#1)}}}
\newcommand{\dchi}[1]{\var{\chi_{(#1)}}}

\newcommand{\phiGI}{Z^{\phi}}
\newcommand{\chiGI}{Z^{\chi}}

\newcommand{\rev}[1]{{#1}}

\begin{document}
\title{Dynamical Stability and Critical Exponents of the Neutral (S-type) Gubser-Rocha Model with Momentum Dissipation}
\author{Shuta Ishigaki}
\affil{%
Department of Physics,
Shanghai University,
99 Shangda Road,
Shanghai 200444, China
}

\maketitle

\begin{abstract}
The Gubser-Rocha model is a holographic model that shows the linear dependence of the entropy density on the temperature.
The model can be extended to possess S-duality by introducing an axio-dilaton field.
With an appropriate choice of the boundary action, both models exhibit a continuous phase transition in the neutral limit.
In this paper, we investigate several aspects of this phase transition.
First, we show that the critical exponents of the phase transition match those of the mean-field percolation theory.
Next, we analyze the dynamical stability, and the emergence of the Nambu-Goldstone modes by analyzing the quasinormal modes of the perturbation fields. The dynamical stability is consistent with the thermodynamic stability. In addition, we find that there is an emergent Nambu-Goldstone mode in the broken phase of the S-type model.
\end{abstract}

\section{Introduction}

The Gubser-Rocha model is known as a holographic model showing linear-$T$ dependence of the entropy density \cite{Gubser:2009qt}.
The four-dimensional model can be obtained from the maximal $\mathcal{N}=8$, $SO(8)$ gauged supergravity, and the eleven-dimensional supergravity \cite{Cvetic:1999xp, Chong:2004na, Chow:2013gba}.
The corresponding objects in the M-theory are considered as particular M2-branes with angular momenta in the extra dimensions.
By introducing massless pseudo-scalar fields that break translational invariance \cite{Andrade:2013gsa}, one can compute the finite DC conductivity.%
\footnote{
	It differs from the finite part of the conductivity studied, e.g., in \cite{Alishahiha:2012ad}.
	Such a finite contribution, which exists when the system has translational invariance, is called incoherent conductivity.
}
Then, the Gubser-Rocha model also reproduce the linear-$T$ dependence of the resistivity \cite{Jeong:2018tua}.
The linear-$T$ resistivity is one of the typical behaviors of strange metals, which is a common non-Fermi liquid phase in hole-doped high-temperature superconductors.
In this context, the various aspects of the Gubser-Rocha model has been studied, e.g., \cite{Zhou:2015qui, Caldarelli:2016nni, Jeong:2021wiu, Wang:2023rca, Ahn:2025tjp}.
The quadratic-$T$ dependence of the inverse Hall angle is also expected for strange metals, but the Gubser-Rocha model does not reproduce it \cite{Ahn:2023ciq}.

In \cite{Ge:2023yom}, they introduced an extended model of the Gubser-Rocha model for the purpose of studying the strange metallic behavior.
Their model involves an additional axion field for that the model enjoys an S-dual transformation, so we refer this model as the S-type Gubser-Rocha model.
The dilaton and the axion field can be treated as a complex scalar field we call axio-dilaton collectively.
By virtue of the S-duality, they obtained an analytic solution of the dyonic black hole solution in this model.
Although the S-type Gubser-Rocha model also does not reproduce the strange metallic behavior of the inverse Hall angle, the presence of the analytic black hole solution allows us to compute analytic expressions of several quantities.
In \cite{Ishigaki:2024djz}, we have determined an appropriate choice of the boundary action and computed thermodynamic potentials of the S-type Gubser-Rocha model.
One of the nontrivial results is that the S-type model has a global $SO(2) \simeq U(1)$ symmetry in the bulk, which is a subgroup of $SL(2,\mathbb{R})$ symmetry for the kinetic term of the axio-dilaton field, in the neutral limit.
Remarkably, the \rev{axio-dilaton part} of the S-type model coincides with a special case of the $U(1)^4$ gauged supergravity model.
The bulk global symmetry is spontaneously broken at a critical temperature determined by the momentum dissipation scale.
The condensation is analytically given by the leading coefficient of the axio-dilaton field at the boundary, and it linearly vanishes at the critical temperature.

In this paper we investigate some aspects of the phase transition of the S-type Gubser-Rocha model with the momentum dissipation.
Firstly, we investigate some of the static critical exponents for this phase transition.
The phase transition itself has been known in the original Gubser-Rocha model, e.g., \cite{Caldarelli:2016nni}, and in the S-type model \cite{Ishigaki:2024djz}.
From the analytic results of the condensate and the heat capacity, we obtain two critical exponents $\beta = 1$ and $\alpha = -1$, respectively.
By using scaling relations, we also obtain the other exponents.
The exponent $\gamma$ associated with the susceptibility is also checked by the numerical computation.
The result of the critical exponents deviate from the usual mean field values, which are also typical values of the usual holographic superfluid model.\cite{Maeda:2009wv}%
\footnote{
	Some holographic models of non-equilibrium steady states also show the mean field values \cite{Nakamura:2012ae, Matsumoto:2018ukk, Zeng:2018ero}.
}
Our result rather agrees with the values of the mean-field percolation theory.
The mean-field percolation theory can be described by the Ginzburg-Landau theory with a cubic potential, and it may be related to taking the triple-trace deformation in our model.

Secondly, we analyze the quasinormal modes (QNMs) of the perturbation fields to investigate the dynamical (in)stability and the emergence of the Nambu-Goldstone (NG) modes associated with the $U(1)$ symmetry breaking in the bulk.
We expect that the axio-dilaton perturbation is related to those behaviors.
By taking into account the diffeomorphism gauge symmetry, we analyze the QNMs of the relevant perturbations.
For the amplitude (Higgs) modes, the behavior of the quasinormal frequency reflects the phase transition at the critical temperature correctly.
Thus, at least within our analysis, the thermodynamic stability agrees with the dynamical stability.
We also find the emergence of the massless NG modes in the symmetry broken phase.
The NG mode is dissipative, and type A following to the categorization in \cite{Minami:2015uzo}.
When the temperature is close to the critical temperature, the dispersion relation for the NG modes is well-fitted by a simple formula up to quadratic order of the frequency and momentum.

This paper is organized as follows.
In section \ref{sec:setup}, we review the S-type Gubser-Rocha model and its solution studied in \cite{Ge:2023yom,Ishigaki:2024djz}.
In section \ref{sec:exponents}, we show the values of the critical exponents for the phase transition in the neutral limit.
In section \ref{sec:perturbation}, we consider perturbation fields to investigate the dynamical stability of the background solution, and the emergence of the Nambu-Goldstone modes associated with the symmetry breaking.
We present the conclusions and the discussions in section \ref{sec:discussion}.

\section{S-type Gubser-Rocha model}\label{sec:setup}
In this section, we briefly review the properties of the four-dimensional S-type Gubser-Rocha model \cite{Ge:2023yom,Ishigaki:2024djz}.
The first part of the action is a four-dimensional Einstein-Maxwell-axion-dilaton theory, whose Lagrangian density is given by
\begin{equation}\label{eq:S-completed_tau}
	\frac{1}{\sqrt{-g}} \mathcal{L}_1
	=
	R - \frac{3}{2}\frac{\partial_{\mu}\tau\partial^{\mu}\bar{\tau}}{(\Im\tau)^2}
	- \frac{1}{4}e^{-\phi} F^2 + \frac{1}{4} \chi F \tilde{F} + \frac{3}{L^2} \frac{\tau\bar{\tau} + 1}{\Im \tau},
\end{equation}
where $\tau = \chi + i e^{-\phi}$ is a complex axio-dilaton field, and $\tilde{F}^{\mu\nu} = \frac{1}{2 \sqrt{-g}} \epsilon^{\mu\nu\rho\sigma} F_{\rho\sigma}$.
In terms of  $\chi$ and $\phi$, we write
\begin{equation}\label{eq:S-completed}
	\frac{1}{\sqrt{-g}} \mathcal{L}_1
	=
	R - \frac{3}{2} (\partial \phi)^2 - \frac{3}{2} e^{2\phi} (\partial\chi)^2 
	-\frac{1}{4} e^{-\phi} F^2 + \frac{1}{4} \chi F \tilde{F} + \frac{1}{L^2}(6\cosh\phi + 3\chi^2 e^{\phi}).
\end{equation}
The model can be related to the $\mathcal{N}=2$, $U(1)^4$ gauged supergravity except the gauge field terms; See Appendix \ref{appendix:sugra} for more details.
The action is invariant under the following transformation:
\begin{align}
     \begin{split}
          \tau &\to  \tau' =\frac{\tau\cos\theta+\sin\theta}{-\tau\sin\theta+\cos\theta},\\
          F &\to  F' = (-\chi\sin\theta+\cos\theta) F -\sin\theta\, e^{-\phi} \tilde{F},\\
          \tilde{F} &\to \tilde{F} ' = (-\chi\sin\theta +\cos\theta) \tilde{F} +\sin\theta\, e^{-\phi} F.
     \end{split}
     \label{eq:U(1)_symmetry}
\end{align}
The transformation for $\tau$ is a subgroup of $SL(2,\mathbb{R})$ transformation, which preserves the kinetic term of $\tau$.%
\footnote{
	One can also write the action in another form having the $U(1)$ symmetry explicitly by redefining the axio-dilaton field. See Appendix \ref{appendix:another_action}.
}
For $\theta = \pi/2$, the transformation is reduced to the S-dual transformation.
The second part of the Lagrangian density is given by
\begin{equation}
	\frac{1}{\sqrt{-g}} \mathcal{L}_{2} = -\frac{1}{2}\sum_{I = 1,2} (\partial \psi_{I})^2.
\end{equation}
With the ansatz we will set below, the massless pseudoscalar fields break the translational symmetry explicitly.
The total Lagrangian density is given by $\mathcal{L} = \mathcal{L}_{1} + \mathcal{L}_{2}$.

The model admits the following dyonic black-brane solution, which was obtained in \cite{Ge:2023yom} analytically.
For the metric ansatz
\begin{equation}
	\dd{s}^2 = - f(r) \dd{t}^2 + \frac{\dd{r}^2}{f(r)} + h(r)(\dd{x}^2 + \dd{y}^2),
\end{equation}
the solution is obtained as
\begin{equation}\label{eq:solution_1}
\begin{gathered}
	f(r) = h(r)\left(
		\frac{1}{L^2} - \frac{n^2 + B^2}{3 \bb(\bb + r)^3} - \frac{\mm^2 \bb}{\bb(\bb + r)^2}
	\right),\quad
	h(r) = \sqrt{r(r+\bb)^3},\\
	e^{-\phi} = \frac{
		(n^2 + B^2)\sqrt{r(r+\bb)}
	}{
		(n^2 + B^2)r + B^2 \bb
	},\quad
	\chi = - \frac{B n \bb}{(n^2 + B^2) r + B^2 \bb},\\
	A = n \left(\frac{1}{r_0 + \bb} - \frac{1}{r + \bb}\right)\dd{t} - \frac{B}{2}y \dd{x} + \frac{B}{2}x \dd{y},\quad
	\psi_{I} = (\sqrt{2} \mm x, \sqrt{2} \mm y),
\end{gathered}
\end{equation}
where $B$ is an external magnetic field which is orthogonal to the $x$--$y$ plane, and $\mm$ is a strength of the momentum relaxation.
The factor of $\sqrt{2}$ in $\psi_{I}$ is taken as our convention.
The black hole horizon is located at $r=r_0$ and the AdS boundary at $r=\infty$.
The real parameter $\bb$ determines the scale of the curvature singularity at $r=-\bb$.
The charge density $n$ is related to the other parameters by
\begin{equation}\label{eq:density}
	n = \sqrt{3\bb(r_0 + \bb)^3\left(
		1 - \frac{B^2}{3\bb(r_0 + \bb)^3} - \frac{\mm^2}{(r_0 + \bb)^2}
	\right)}.
\end{equation}
The chemical potential is given by $\mu = n/(r_0 + \bb)$.
In this paper, we also employ the following parametrization:
\begin{equation}\label{eq:def_Q_theta}
	Q:= \sqrt{n^2 + B^2}
	= \sqrt{3\bb(r_0 + \bb)^3\left(1 -  \frac{\mm^2}{(r_0 + \bb)^2}\right)},\quad
	\frac{\theta}{2} := \atan\frac{B}{n}.
\end{equation}
In other words, we write $n = Q \cos\frac{\theta}{2}$ and $B= Q \sin\frac{\theta}{2}$.
It is noteworthy that the following combination gives a $r$-constant:
\begin{equation}
	\frac{2 \chi}{1 - e^{-2\phi} - \chi^2} = \frac{2 B n}{n^2 - B^2}
	= \tan\theta.
\end{equation}
The Hawking temperature and entropy density are obtained as
\begin{equation}
	T = \frac{r_0}{4\pi\sqrt{r_0(r_0 + \bb)^3}}\left(
		3(r_0+\bb)^2 - \mm^2
	\right),\quad
	s = 4\pi\sqrt{r_0(r_0 + \bb)^3},
\end{equation}
respectively.
The family of solution is parameterized by $(r_0, \bb, \mm, B)$ corresponding to the boundary quantities $(T, \mu, \mm, B)$.
Hereinafter, we set $L=1$.

To read the coefficients of the asymptotic expansions of each field, we need to fix the radial coordinate to the Fefferman-Graham coordinate $z$, whose metric has a form of
\begin{equation}
	\dd{s}^2 = \frac{h_{ij}(z) \dd{x^{i}} \dd{x^{j}} + \dd{z}^2}{z^2}.
\end{equation}
The coordinate transformation from $r$ to $z$ can be asymptotically written as
\begin{equation}
	r = 
	\frac{1}{z}
	-\frac{3 \bb}{4}
	+ \left(\frac{3 \bb^2}{64} + \frac{\mm^2}{4}\right) z
	+\frac{\left(3 \bb^4-24 \mm ^2 \bb^2 + 16 Q^2\right)}{288 \bb }z^2 
	+\order{z^3}.
\end{equation}
The nonzero components of the metric are expanded as
\begin{subequations}\label{eq:hij_FG_expansion}
\begin{align}
	h_{tt} =& - 1
	+ \left(
		\frac{3 \bb^2}{32} + \frac{\mm^2}{2}
	\right)z^2
	+ \left(
		\frac{\bb^3}{24} + \frac{2 Q^2}{9 \bb} - \frac{\bb \mm^2}{3}
	\right)z^3 + \order{z^4},\\
	h_{xx} =& 1
	+ \left(
		- \frac{3 \bb^2}{32} + \frac{L^2 \mm^2}{2}
	\right) z^2
	+ \left(
		- \frac{\bb^3}{24} + \frac{Q^2}{9 \bb} - \frac{\bb \mm^2}{6}
	\right) z^3 + \order{z^4},
\end{align}
\end{subequations}
and $h_{yy}=h_{xx}$.
We write the expansion of $h_{ij}$ in the vicinity of $z=0$ as
\begin{equation}
	h_{ij} = h_{ij}^{(0)} + h_{ij}^{(1)} z + h_{ij}^{(2)} z^2 + h_{ij}^{(3)} z^3 + \order{z^4}.
\end{equation}
The expansions for the dilaton and axions become
\begin{align}
	\phi =& - \frac{\bb}{2} \cos\theta z
	- \frac{\bb^2}{16}\left(
		2 \cos\theta + \cos2\theta -1
	\right) z^2 + \order{z^3},\\
	\chi =&
	\frac{\bb}{2} \sin\theta z
	+
	\frac{\bb^2}{8}\left(
		\sin\theta +\sin2\theta
	\right) z^2 + \order{z^3},
\end{align}
respectively.
In the same manner, we also write the coefficients as
\begin{align}
	\phi = \phi_{(1)} z + \phi_{(2)} z^2 + \order{z^3},\\
	\chi = \chi_{(1)} z + \chi_{(2)} z^2 + \order{z^3}.
\end{align}


In \cite{Ishigaki:2024djz}, we have determined the counterterm and the boundary action associated with the above solution, in which the boundary theory becomes conformal, and source-free for the operator dual to the axio-dilaton.
The boundary action is deduced from one obtained in \cite{Caldarelli:2016nni} by generalizing it to the axio-dilaton case.
\rev{%
In the boundary field theory, it corresponds to taking multi-trace deformations.\cite{Witten:2001ua}
}%
With a small cutoff $\epsilon$, the counterterm is given by
\begin{equation}\label{eq:Sct_1}
\begin{aligned}
	S_{\text{ct}} =
	- \int_{z=\epsilon} \sqrt{-\gamma}\left[
		4 - \frac{1}{2}\sum_{I} \gamma^{ij}\partial_{i}\psi_{I}\partial_{j}\psi_{J}
		+ \frac{3}{2} \phi^2 + \frac{3}{2} e^{\phi} \chi^2
	\right],
\end{aligned}
\end{equation}
and the boundary action
\begin{equation}\label{eq:Sfin}
\begin{gathered}
	S_{\text{fin}} = \int\dd[3]{x}\sqrt{-h_{(0)}}\left[
		\phi_{(1)} J_{\phi} + \chi_{(1)} J_{\chi} + \mathcal{F}
	\right],\\
	J_{\phi} := - 3\left(\phi_{(2)} - \frac{1}{2}\chi_{(1)}^2\right)
	- \pdv{\mathcal{F}}{\phi_{(1)}},\quad
	J_{\chi} := - 3\left(\chi_{(2)} + \phi_{(1)} \chi_{(1)}\right)
	- \pdv{\mathcal{F}}{\chi_{(1)}}.
\end{gathered}
\end{equation}
The function $\mathcal{F}$ determines the way of the deformation in the boundary theory.
We set
\begin{equation}\label{eq:deformation_F}
	\mathcal{F} := - \frac{1}{2}\left(\phi_{(1)}^2 + \chi_{(1)}^2\right)^{3/2},
\end{equation}
\rev{%
which implies considering a triple-trace deformation.
}%
The total action is given by
\begin{equation}
	S_{\rm ren} := \lim_{\epsilon\to0} S_{\rm reg} + S_{\rm GH} + S_{\rm ct} + S_{\rm fin},
\end{equation}
where $S_{\rm reg}$ is the regularized bulk action, and $S_{\rm GH}$ is the Gibbons-Hawking action; see \cite{Ishigaki:2024djz} for details.
From the variation of the total action, we obtain the expressions of the sources as
\begin{equation}\label{eq:sources}
\begin{aligned}
	\frac{J_{\phi}}{3} =& -\left(
		\phi_{(2)} - \frac{1}{2} \chi_{(1)}^2
	\right)
	+ \frac{1}{2} \phi_{(1)}
	\sqrt{\phi_{(1)}^2 + \chi_{(1)}^2},\\
	\frac{J_{\chi}}{3} =& -\left(
		\chi_{(2)} + \phi_{(1)}\chi_{(1)}
	\right)
	+ \frac{1}{2} \chi_{(1)}
	\sqrt{\phi_{(1)}^2 + \chi_{(1)}^2}.
\end{aligned}
\end{equation}
and the operators as
\begin{equation}
	O_{\phi} := \phi_{(1)},\quad
	O_{\chi} := \chi_{(1)}.
\end{equation}
The black hole solution (\ref{eq:solution_1}) satisfies $J_{\phi}=J_{\chi}=0$ only when $\bb \geq 0$.%
\footnote{
	If one inserts $\mathrm{sgn}(\bb)$ into $\mathcal{F}$, the solution with $\bb < 0$ also satisfies the source-less condition.
}
We also define
\begin{equation}
	O := O_{\phi} + i O_{\chi}.
\end{equation}
With the black hole solution, we obtain $O = - \frac{b}{2} \exp({-i\theta})$.
In this quantization, the boundary stress tensor can be obtained as
\begin{equation}
	\mathcal{T}_{ij} := 3 h_{ij}^{(3)} + \left(
		\phi_{(1)}^2 + \chi_{(1)}^2
	\right)^{3/2} h_{ij}^{(0)}.
\end{equation}
Using the expansions (\ref{eq:hij_FG_expansion}), we obtain
\begin{equation}
	\mathcal{T}_{tt} = 2 \mathcal{T}_{xx} = 2 \mathcal{T}_{yy}
	= \frac{2 Q^2}{3 \bb} - \bb m^2.
\end{equation}
Thus one obtains $\mathcal{T}^{i}{}_{i}=0$.
From the on-shell action, the grand potential density is also determined as
\begin{equation}\label{eq:grand_potential}
	\Omega := - (r_{0} + \bb)^3 - \mm^2 r_{0} + \frac{B^2}{r_{0} + \bb}.
\end{equation}

\subsection{Phase transition in the neutral limit}
The S-type Gubser-Rocha model exhibits spontaneous $U(1)$ symmetry breaking in the neutral limit.
If we ignore the presence of the axion field, this is simply reduced to $Z_{2}$ breaking in the original Gubser-Rocha model.
From Eq.~(\ref{eq:def_Q_theta}), we can take $Q=0$ by the following two ways:
\begin{equation}
	\bb = 0\quad
	\text{or}\quad
	\bb = \mm - r_{0}.
\end{equation}
Applying these conditions to the solution, we obtain $\tau=i$ for $\bb=0$ and $\tau\neq i$ for $\bb = \mm - r_{0}$, respectively.
Both solutions can take the temperature range in $2\pi T/\mm \in [0, \infty)$.
The ground state is determined by comparing the thermodynamic potential (\ref{eq:grand_potential}) between them.
As a result, the solution with $\tau\neq i$, $\bb=\mm - r_{0}$, is the ground state for $0 \leq 2\pi T/\mm \leq 1$ whereas the solution with $\tau = i$, $\bb=0$, is for $1 < 2\pi T/\mm$.
The critical point, $b=0$ and $\mm =r_{0}$, is a special point that an additional symmetry is emerged in the equations of motion.\cite{Davison:2014lua}

\begin{figure}[htbp]
\centering
\includegraphics[width=0.49\linewidth]{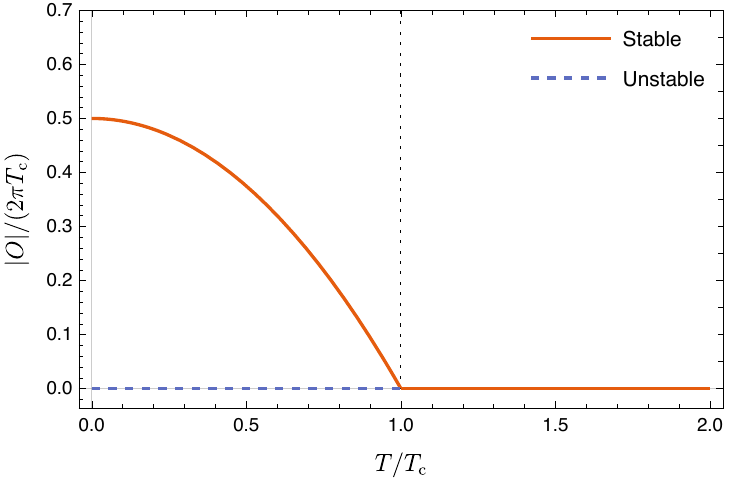}
\includegraphics[width=0.49\linewidth]{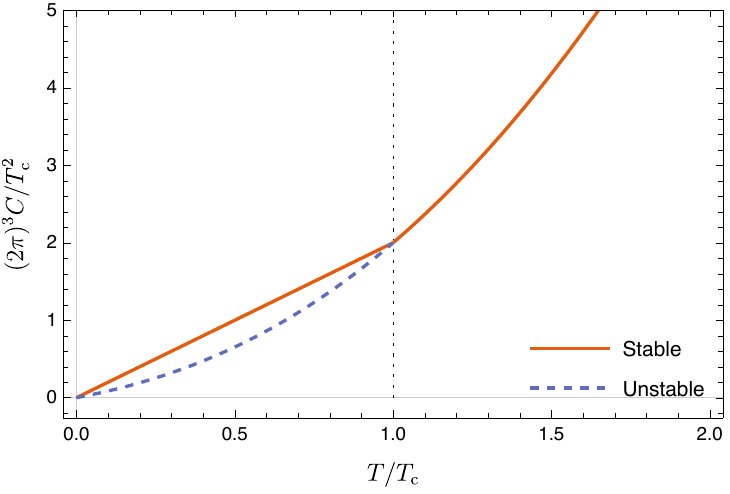}
\caption{
	Left: Vacuum expectation value of the dual operator to the axio-dilaton as a function of temperature.
	Right: Heat capacity as a function of temperature.
}
\label{fig:condensate}
\end{figure}

This phase transition is related to the symmetry given by Eq.~(\ref{eq:U(1)_symmetry}).
Note that the transformation (\ref{eq:U(1)_symmetry}) becomes symmetry of the action only when $F$ can be ignored in the neutral limit.
The solution with $\tau=i$, $\bb=0$, corresponds to the restored phase, whereas the solution with $\tau \neq i$, $\bb = \mm - r_{0}$, the broken phase.
The order parameter is given by the dual operator to the axio-dilaton field, which is expressed as
\begin{equation}\label{eq:condensate}
	O = \phi_{(1)} + i \chi_{(1)}
	=
	- \pi T_{\rm c} \left(
		1 - \frac{T^2}{T_{\rm c}^2}
	\right) e^{i\theta},
\end{equation}
for $0\leq T< T_{\rm c}$, and $O=0$ for $T_{\rm c} \leq T$, where $T_{c}:= m/2\pi$.
\rev{%
The result suggests that the phase transition is second order since $\pdv{O}{T} = \pdv{\Omega}{T}{J}$ shows a discontinuity at $T=T_{\rm c}$.
}%
We can also compute the heat capacity as
\begin{equation}\label{eq:heat_capacity}
	C := - T \pdv[2]{\Omega}{T}
	=
	\frac{T_{\rm c}^2}{(2\pi)^3}\times
	\begin{cases}
		\frac{T}{T_{\rm c}} & 0\leq T<T_{\rm c}\\
		\frac{8}{9}\frac{T^2}{T_{\rm c}^2}
		+\frac{1}{3 \sqrt{3 + T^2/T_{\rm c}^2}}\left(
			\frac{4}{3}\frac{T}{T_{\rm c}}
			+ \frac{8}{9}\frac{T^3}{T_{\rm c}^3}
		\right)
		& T_{\rm c} \leq T
	\end{cases}.
\end{equation}
The expansion around $T=T_{\rm c}$ becomes
\begin{equation}
	(2\pi)^3 \frac{C}{T_{\rm c}^2} = 2 +
	\begin{cases}
		2 (\frac{T}{T_{\rm c}} - 1) + \order{\frac{T}{T_{\rm c}} - 1}^2
		& 0 \leq T < T_{\rm c}\\
		\frac{7}{2} (\frac{T}{T_{\rm c}} - 1) + \order{\frac{T}{T_{\rm c}} - 1}^2
		& T_{\rm c} \leq T
	\end{cases}.
\end{equation}
\rev{%
The result suggests that $\Omega$ itself shows a discontinuity from its third derivative with respect to temperature, similar to \cite{Ren:2019lgw}.
However, as suggested by the behavior of $O$, the phase transition should be classified into the second order one.
}%
We show the behaviors of the condensate and the heat capacity as functions of temperature in figure \ref{fig:condensate}.

\section{Critical exponents}\label{sec:exponents}
Critical exponents characterize a continuous phase transition in general.
Let us check the critical exponents of the phase transition in the neutral limit in our model.
The (static) critical exponents are defined by the following scaling behaviors:
\begin{subequations}
\begin{gather}
	C \sim \abs{1 - \frac{T}{T_{c}}}^{-\alpha}, \quad (T<T_{c})\\
	\expval{\mathcal{O}} \sim \abs{
		\frac{T}{T_{c}} - 1
	}^{\beta},\quad (T>T_{c})\\
	\left.\pdv{\expval{\mathcal{O}}}{J}\right|_{J=0}
	\sim \abs{
		1 - \frac{T}{T_{c}}
	}^{-\gamma},\quad (T>T_{c})\\
	J \sim \expval{\mathcal{O}}^{\delta},\quad (T = T_{c})\\
	\expval{\mathcal{O}(\vec{x})\mathcal{O}(0)}
	\sim e^{-|x|/\xi}
	\sim \abs{x}^{-d_{\rm s} +2-\eta},\quad (T=T_{c})\\
	\xi \sim
	\abs{
			1 - \frac{T}{T_{c}}
	}^{-\nu}.
\end{gather}
\end{subequations}
where $C$ is a heat capacity, $\mathcal{O}$ is an operator of the order parameter, and $J$ is a source.
$d_{\rm s}$ is a spatial dimension.
$\alpha'$ and $\gamma'$ are also defined for $T<T_{c}$ in the same manner.
$\xi$ is a correlation length associated with the two point function of $\mathcal{O}$.

In our case, we regard $O$ as the expectation value of the operator $\mathcal{O}$.
From Eqs.~(\ref{eq:condensate}) and (\ref{eq:heat_capacity}), we obtain $\beta = 1$ and $\alpha = -1$.%
\footnote{
	Although the expansion of $C$ around $T=T_{c}$ begins from the zeroth order term, we read $\alpha=-1$ since $C$ is continuous at $T=T_{c}$.
	The case of $\alpha=0$ usually means presence of a discontinuous jump at $T=T_{c}$.
}
It is expected that there are scaling relations
\begin{subequations}\label{eq:scaling_relations}
\begin{gather}
	\alpha + 2\beta + \gamma = 2,\label{eq:first_scaling_relation}\\
	\gamma = \beta (\delta - 1),\\
	\gamma = \nu (2-\eta),\\
	2-\alpha = \nu d_{\rm s},
\end{gather}
\end{subequations}
where $d_{\rm s} = 2$ in our case.
The last line including $d_{\rm s}$ is called hyperscaling relation.
If we use the above relations, only two exponents are independent actually.
As a result, we obtain $\gamma =1$ and $\delta = 2$.
The result $\delta = 2$ is also consistent with the scaling dimensions of the operator and source: $[O_{\phi}]=1$ and $[J_{\phi}]=2$.
If we use the hyperscaling relation, we also obtain $\nu = 3/2$ and $\eta = 4/3$.
Later, we will check the value of $\gamma$ from the Green's function numerically.
In summary, we obtain the critical exponents of the $U(1)$ symmetry breaking as
\begin{equation}\label{eq:critical_exponents}
	\alpha = \alpha' = -1,\quad
	\beta = 1,\quad
	\gamma = \gamma' = 1,\quad
	\delta = 2,\quad
	\left(
	\nu = \frac{3}{2},\quad
	\eta = \frac{4}{3},
	\right)
\end{equation}
the last two are obtained by using the hyperscaling relation.
As we will discuss in section \ref{sec:discussion}, the above values of the critical exponents coincide with those in the mean-field percolation theory.

\section{Linear perturbations}\label{sec:perturbation}
To analyze the amplitude modes and the Nambu-Goldstone modes associated with the $U(1)$ symmetry in the neutral limit, we consider small perturbation fields around the background solution.
The frequency of the amplitude mode reflects the dynamical (in)stability of the system.
The amplitude and NG modes are mainly related to the axio-dilaton perturbation fields in our case.
Since the gravitational perturbation fields are gauged under the diffeomorphism, which corresponds to the coordinate transformation invariance, we have to solve the system with the gauge symmetry.
The number of the degrees of freedom in our model is summarized as table \ref{tab:dof}.

\begin{table}
	\centering
	\begin{tabular}{c| c c c c}
		fields & $g_{\mu\nu}$ & $A_{\mu}$ & $\tau$ & $\psi_{I}$\\\hline
		\rev{components} & 10 & 4 & 2 & 2\\
		gauge freedom & 4 & 1 & & \\
		off-shell dof & 6 & 3 & 2 & 2\\
		on-shell dof & 2 & 2 & 2 & 2
	\end{tabular}
	\caption{
		Field contents and their degrees of freedom in our model.
		\rev{%
		The first row shows the number of the field components $N$.
		The second row shows the number of gauge freedoms $G$.
		The off-shell and on-shell degrees of freedom are given by $N-G$ and $N-2G$, respectively.
		}%
		The total number of the on-shell degrees of freedom is eight.
	}
	\label{tab:dof}
\end{table}

\subsection{Gauge invariants}\label{appendix:gauge_invariants}
According to \cite{Kovtun:2005ev}, one way to obtain physically meaningful results in the gravitational perturbation analysis is to consider gauge-invariant combinations of the perturbation fields.
A list of gauge-invariant combinations for the original Gubser-Rocha model without applying magnetic fields has been shown in \cite{Jeong:2023ynk}.
In this subsection, we show a complete set of the gauge invariants in our setup with the axio-dilaton and an applying external magnetic field.

We consider perturbation fields of the plane-wave ansatz, such as
\begin{equation}
	\Phi \to \Phi + \var{\Phi}(r) e^{-i\omega t + i k_{x} x},
\end{equation}
where $\Phi$ denotes the background fields and $\var{\Phi}$ the perturbation fields.
The infinitesimal diffeomorphism plays a role of the gauge symmetry for the perturbation fields.
Let us now consider the infinitesimal diffeomorphism generated by
\begin{equation}
	\epsilon^{\mu} = \left[
		\zeta^{t}(r),~
		\zeta^{x}(r),~
		\zeta^{y}(r),~
		\zeta^{u}(r)
	\right]^{\rm T} e^{-i\omega t + ik_{x} x},
\end{equation}
where $\zeta^{\mu}$ are arbitrary functions of $r$, and we take a diagonal metric background given by
\begin{equation}
	\dd{s}^2 =
	g_{tt}\dd{t}^2
	+ g_{xx}\dd{x}^2
	+ g_{yy}\dd{y}^2
	+ g_{rr}\dd{r}^2.
\end{equation}
We assume that $\epsilon^{\mu}$ is of the same order as $\var{\Phi}$.
The infinitesimal coordinate transformation $x^{\mu} \to x^{\mu} - \epsilon^{\mu}$ induces
\begin{equation}
	\mathcal{T} \to
	 \mathcal{T} + \LieD_{\epsilon} \mathcal{T},
\end{equation}
for any tensor field $\mathcal{T}$, where $\LieD_{\epsilon}$ denotes taking Lie derivative, e.g., $\LieD_{\epsilon} g = 2 \nabla_{\mu} \epsilon_{\nu} \dd{x^{\mu}} \dd{x^{\nu}}$ for the metric.
Since $\epsilon^{\mu}$ is of the same order as $\var{\Phi}$, it can be regarded as the gauge transformation for the perturbations $\var{\Phi}$.

One can find gauge-invariant combinations of the perturbation fields.
For the gravitational perturbations, we find
\begin{subequations}\label{eq:sound_shear}
\begin{align}
	Z_{1}:=& \frac{g^{xx}}{\omega^2}\left[
	\omega^2 \var{g_{xx}}
	+ 2 k_{x} \omega \var{g_{t x}}
	+ k_{x}^2 \var{g_{tt}}
	- \left(
		\omega^2 \frac{\partial_{r} g_{xx}}{\partial_{r} g_{yy}}
		+ k_{x}^2 \frac{\partial_{r} g_{tt}}{\partial_{r} g_{yy}}
	\right)\var{g_{yy}}
	\right],\\
	Z_{2} :=& g^{yy}\left[\omega \var{g_{xy}} + k_{x} \var{g_{ty}}\right].
\end{align}
\end{subequations}
Note that, unlike \cite{Kovtun:2005ev}, there is no helicity-2 channel since the dimension of the field theory is 2+1.
Taking a limit of $k_{x}\to0$, $Z_{2}$ is reduced to $\var{g}_{x}^{y}$, which is related to the shear viscosity.%
\footnote{
    Note that explicit breaking of the translational invariance can violate the viscosity bound.\cite{Hartnoll:2016tri}
}

The infinitesimal diffeomorphism also acts on the (pseudo)scalar fields.
For the perturbations of the pseudoscalars breaking the translational invariance, we find the gauge invariants as
\begin{align}
	Z_{3} :=& i k_{x}  \var{\psi_{x}}
	- \frac{\sqrt{2}\mm}{2} g^{xx}\left(
		\var{g_{xx}} - \frac{\partial_{r} g_{xx}}{\partial_{r} g_{yy}} \var{g_{yy}}
	\right),\\
	Z_{4} :=& i k_{x} \var{\psi_{y}} - \sqrt{2} \mm g^{yy} \var{g_{xy}}.
\end{align}
For the dilaton and the axion perturbations, we can find the gauge invariants as
\begin{equation}
	Z_{5} := \var{\phi} - \frac{\partial_{r} \phi}{\partial_{r} g_{yy}} \var{g_{yy}},\quad
	Z_{6} := \var{\chi} - \frac{\partial_{r} \chi}{\partial_{r} g_{yy}} \var{g_{yy}},
\end{equation}
respectively.
Now, we take a linear combination
\begin{equation}
\begin{aligned}
	\phiGI:=& Z_{5}
	- \frac{\partial_{r}\phi}{\partial_{r} g_{xx} + \partial_{r} g_{yy}} g_{xx} Z_{1}\\
	=&
	\var{\phi}
	- \frac{\partial_{r}\phi}{\omega^2(\partial_{r} g_{xx} + \partial_{r} g_{yy})}  \left[
		\omega^2 \var{g_{xx}} + 2 \omega k_{x} \var{g_{tx}} + k_{x}^2 \var{g_{tt}}
		+ \left(
			\omega^2
			- k_{x}^2 \frac{\partial_{r}g_{tt}}{\partial_{r} g_{yy}}
		\right) \var{g_{yy}}
	\right],
\end{aligned}
\end{equation}
and $\chiGI$ for $\chi$ as well.
Such linear combinations of the gauge invariants are also gauge invariants.
In the limit of $k_{x}=0$, $Z_{5}$ and $Z_{6}$ are coupled with $Z_{1}$, but $\phiGI$ and $\chiGI$ are not.
Thus, it is more convenient to use them than $Z_{5}$ and $Z_{6}$ in this case.

The vector fields are also transformed under the electromagnetic $U(1)$ gauge transformation.
Considering
\begin{equation}
	\Lambda = \Lambda(r) e^{-i\omega t + i k_{x} x},
\end{equation}
we obtain the total infinitesimal variation as
\begin{equation}
	\Delta_{\epsilon,\Lambda} \var{A} :=
	\LieD_{\epsilon} A + \dd{\Lambda}.
\end{equation}
Under this transformation, we can also find the gauge invariants as
\begin{align}
	Z_{7} :=&  i \omega \var{A_{x}} + i k_{x} \var{A_{t}}- i k_{x} \frac{\partial_{r} A_{t}}{\partial_{r} g_{yy}} \var{g_{yy}}
	- \frac{B}{2} g^{yy} \var{g_{ty}},\\
	Z_{8} :=& i k_{x} \var{A_{y}}
	- \frac{B}{4} g^{xx}\left(
		\var{g_{xx}}
		- \frac{\partial_{r} g_{xx}}{\partial_{r}g_{yy}} \var{g_{yy}}
	\right).
\end{align}
We have eight gauge invariants $Z_{i}$ corresponding to the on-shell degrees of freedom shown in table \ref{tab:dof}.

\subsection{Boundary conditions for the perturbation fields}
The sources for the operators dual to the axio-dilaton are given by Eq.~(\ref{eq:sources}).
Considering linear perturbations, we obtain sources for the axio-dilaton perturbations as
\begin{align}
	(1/3)\var{J_{\phi}}
	=&
	- \dphi{2}
	+ \frac{1}{8} \bb  (\cos2\theta +3) \dphi{1}
	-\frac{1}{8} \bb  (\sin2\theta - 4 \sin\theta) \dchi{1},\\
	(1/3)\var{J_{\chi}}
	=&
	- \dchi{2}
	- \frac{1}{8} \bb (\sin2\theta + 4 \sin \theta) \dphi{1}
	-\frac{1}{8} \bb  (-4 \cos\theta+\cos2\theta-3) \dchi{1}.
\end{align}
We can also write the above expressions in terms of the gauge invariants:
\begin{align}
	(1/3)\var{J_{\phi}}
	=&
	- \phiGI_{(2)}
	+ \frac{1}{8} \bb  (\cos2\theta +3) \phiGI_{(1)}
	- \frac{1}{8} \bb  (\sin2\theta - 4 \sin\theta) \chiGI_{(1)},\\
	(1/3)\var{J_{\chi}}
	=&
	- \chiGI_{(2)}
	- \frac{1}{8} \bb \sin\theta (\sin2\theta + 4\sin\theta) \phiGI_{(1)}
	-\frac{1}{8} \bb  (-4 \cos\theta+\cos2\theta-3) \chiGI_{(1)},
\end{align}
where $Z_{(k)}^{\phi,\chi}$ are defined by
\begin{align}
	Z^{\phi,\chi} = z Z_{(1)}^{\phi,\chi} + z^2 Z_{(2)}^{\phi,\chi} + \order{z^3},
\end{align}
respectively.
Interestingly, the expression takes the same form to the original one.
One can find the similar result even for using $Z_{5}$ and $Z_{6}$.
To obtain the QNMs, we impose $\var{J_{\phi}} = \var{J_{\chi}} =0$ for $Z_{\phi}$ and $Z_{\chi}$ at the boundary.
At the horizon, we impose the incoming-wave condition for each gauge invariant.
We find such eigen-solutions by the shooting method numerically.
If one needs to compute QNMs with large imaginary frequencies, however, alternative methods such as the pseudospectral method are more suitable.
In the following, we consider only the cases where the perturbation field of interest is decoupled.
Even if the perturbation fields are coupled, we can solve such a problem by using the so-called determinant method \cite{Kaminski:2009dh}.
In this case, we impose vanishing Dirichlet conditions for the other perturbation fields.

\subsection{Dynamical stability and susceptibility}
\begin{figure}[htbp]
\centering
\includegraphics[width=0.49\linewidth]{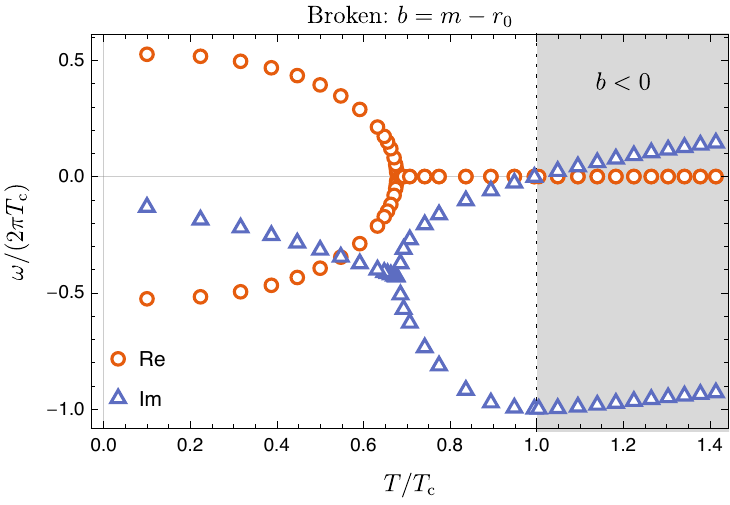}
\includegraphics[width=0.49\linewidth]{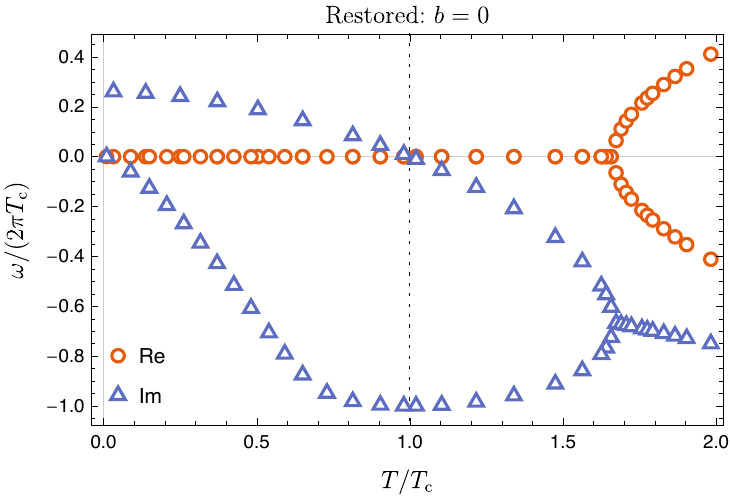}
\caption{
	Quasinormal frequencies of the dilaton (Higgs) modes as a function of temperature.
	Left: the result with $\bb = \mm - r_{0}$.
	The gray region indicates $b<0$ where $J_{\phi}=J_{\chi}=0$ does not hold.
	Right: the result with $\bb = 0$.
}
\label{fig:Higgs_modes}
\end{figure}

In this subsection, we investigate the dynamical stability of the background solution by analyzing the QNMs of $Z_{\phi}$.
For simplicity, we focus on the neutral case.
The perturbation $Z^{\phi}$ is decoupled when $k_{x}=0$ and $\theta =0$.
Since $\theta$ is just an angle of the $U(1)$ symmetry of the background geometry, we can expect that the result does not depend on the specific value of $\theta$.
See also Fig.~\ref{fig:map_xy2uv} in Appendix \ref{appendix:another_action} for the illustration of the $U(1)$ transformation orbits.
In this case, the expression of $Z^{\phi}$ is reduced to
\begin{equation}
	Z^{\phi} = \var{\phi} - \frac{\partial_{r} \phi}{ \partial_{r} g_{xx} + \partial_{r} g_{yy}}
	\left(
		\var{g_{xx}} + \var{g_{yy}}
	\right).
\end{equation}
The linearized equation of motion becomes
\begin{equation}
\begin{aligned}
0 =&
Z_{\phi}''(r) + 
Z_{\phi}'(r)\left(
	\frac{f'(r)}{f(r)} + \frac{h'(r)}{h(r)}
\right)\\
+& Z_{\phi}(r)\left(
	-\frac{6 h(r) \phi '(r)^2 \left(\mm^2-3 h(r) \cosh\phi \right)}{f(r) h'(r)^2}
	+\frac{12 h(r) \phi '(r) \sinh\phi}{f(r) h'(r)}
	+\frac{2 \cosh\phi}{f(r)}
	+\frac{\omega ^2}{f(r)^2}
\right).
\end{aligned}
\end{equation}
Figure \ref{fig:Higgs_modes} shows the behavior of the quasinormal frequencies as functions of the temperature.
From the thermodynamic analysis the broken-phase solution given by $b=m - r_{0}$ is stable at the lower temperatures $T\leq T_{\rm c}$, whereas the restored-phase solution given by $b=0$ at $T>T_{\rm c}$.
We observe that the lowest mode frequency approaches $\omega = 0$ at $T = T_{\rm c}$ in both case.
The mode becomes tachyonic at the side of $T>T_{\rm c}$ for $b=m-r_{0}$, whereas $T<T_{\rm c}$ for $b=0$.
Those behaviors correctly reflect the phase transition at $T = T_{\rm c} = \mm/2\pi$.
We do not observe any other instability in this sector within our analysis.

\begin{figure}[htbp]
\centering
\includegraphics[width=0.49\linewidth]{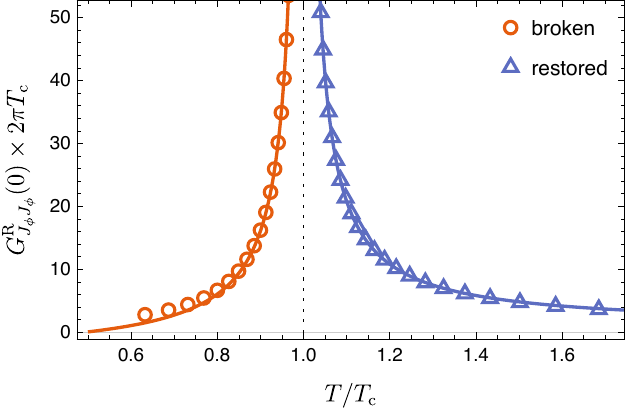}
\includegraphics[width=0.49\linewidth]{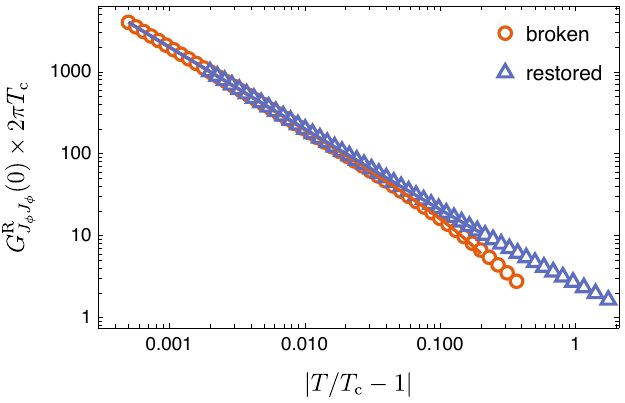}
\caption{
	Left: Susceptibility as a function of temperature.
    The points denote numerical results, and the curves denote fitting with Eq.~(\ref{eq:susceptibility_fit}).
    Right: The same data shown on a double-logarithmic scale with $|T/T_{\rm c} - 1|$ as the x-axis.
}
\label{fig:susceptibility}
\end{figure}

The perturbation $Z_{\phi}$ is also related to the susceptibility.
Let us compute the susceptibility to check the critical exponent $\gamma$ numerically.
We expect that the Green's function can be computed by
\begin{equation}
	G_{J_{\phi} J_{\phi}}^{\rm R}(\omega)
	=
	\left.
	\pdv{\expval{O_{\phi}}}{J_{\phi}}
	\right|_{J_{\phi} = 0}
	=
	\frac{\phiGI_{(1)}}{\var{J_{\phi}}}.
\end{equation}
Taking a limit of $\omega = 0$, we obtain the susceptibility related to the critical exponent $\gamma$.
Figure \ref{fig:susceptibility} shows the behavior of the susceptibility as a function of $T$ in the neutral limit.
The result is well-fitted with
\begin{equation}\label{eq:susceptibility_fit}
    G_{J_{\phi} J_{\phi}}^{\rm R}(0) \simeq
	\frac{C_{\chi}}{|T/T_{\rm c} - 1|} + \chi_{0},
\end{equation}
for $T <  T_{\rm c} <$ and $T_{\rm c} < T$ separately.
$C_{\chi}$, $\chi_{0}$ are fitting parameters, given by
$(C_{\chi} \mm, \chi_{0} \mm) = (1.99967, -3.96189)$ for $T< T_{\rm c}$ and $(C_{\chi} \mm, \chi_{0} \mm) = (2.00038, 0.811002)$ for $T>T_{\rm c}$.
The result of the susceptibility implies $\gamma=\gamma'=1$, which agrees with the previous result (\ref{eq:critical_exponents}).

\subsection{Nambu-Goldstone modes}
The NG mode is related to the axion perturbation $Z^{\chi}$ at $\theta=0$.
In this case, the gauge invariant is simply given by $Z^{\chi} = \var{\chi}$, and it is decoupled even for finite $k_{x}$.
The linearized equation of motion is given by
\begin{equation}
0 = Z_{\chi}''(r) +
Z_{\chi}'(r) \left(\frac{f'(r)}{f(r)}+\frac{h'(r)}{h(r)}+2 \phi '(r)\right)
+Z_{\chi}(r)\left(
	\frac{2}{f(r)} e^{-\phi}
	-\frac{k_{x}^2}{f(r)h(r)}
	+\frac{\omega ^2}{f(r)^2}
\right).
\end{equation}

\begin{figure}[htbp]
\centering
\includegraphics[width=0.49\linewidth]{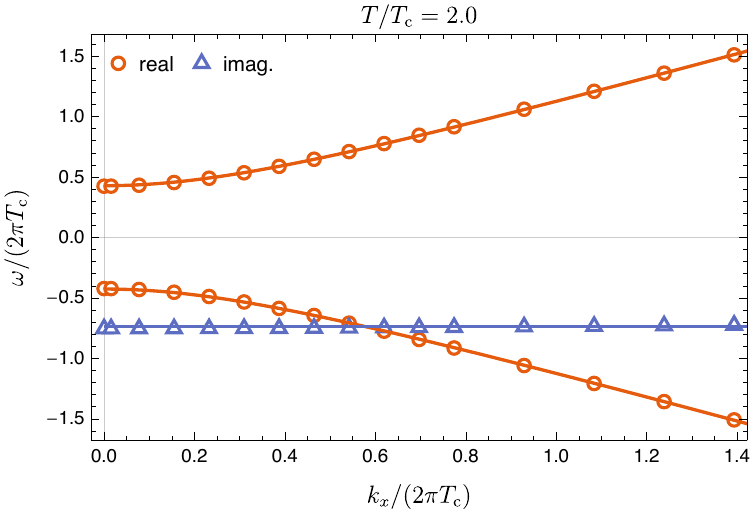}
\includegraphics[width=0.49\linewidth]{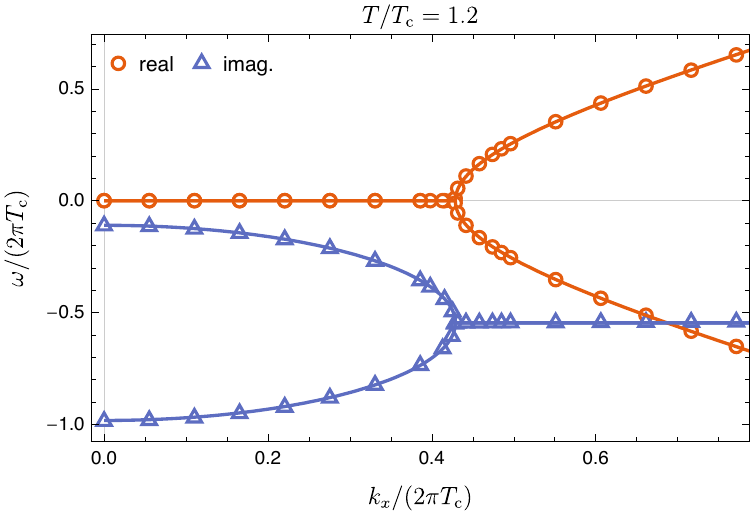}
\includegraphics[width=0.49\linewidth]{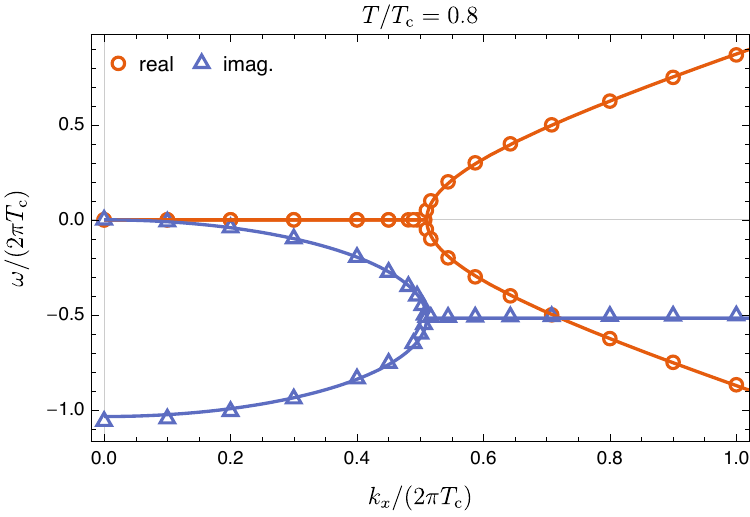}
\caption{
	Dispersion relations of the axion (NG) modes for $\theta=0$.
    The points denote the numerical results, while the curves denote fitting results with Eq.~(\ref{eq:dispersion_relation}).
	Top-left: restored phase at $T/T_{\rm c}= 2.0$.
	Top-right: restored phase at $T/T_{\rm c} = 1.2$.
	Bottom: broken phase at $T/T_{\rm c} = 0.8$.
}
\label{fig:NG_modes}
\end{figure}

\begin{figure}[htbp]
\centering
\includegraphics[width=0.49\linewidth]{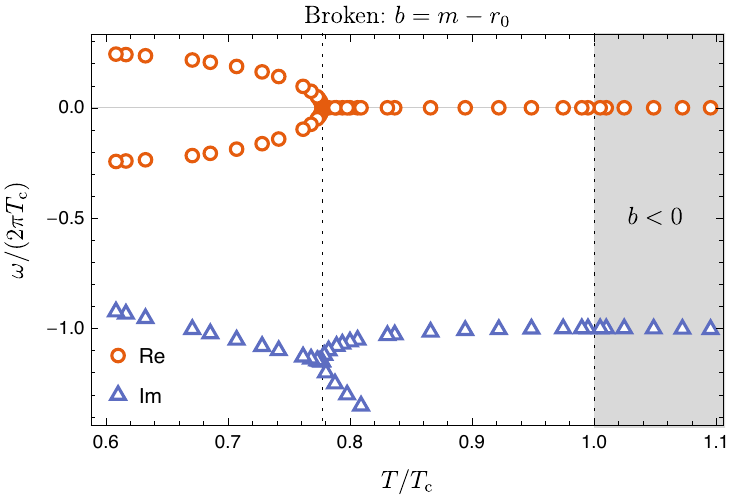}
\caption{
	Quasinormal frequency of the second NG mode as a function of the temperature.
	The vertical dotted line shows $T=0.77718 T_{\rm c}$ and $T=T_{\rm c}$.
}
\label{fig:second_NGmode}
\end{figure}

\begin{figure}[htbp]
\centering
\includegraphics[width=0.33\linewidth]{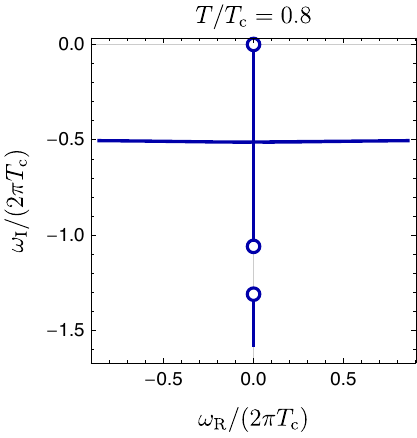}\quad
\includegraphics[width=0.33\linewidth]{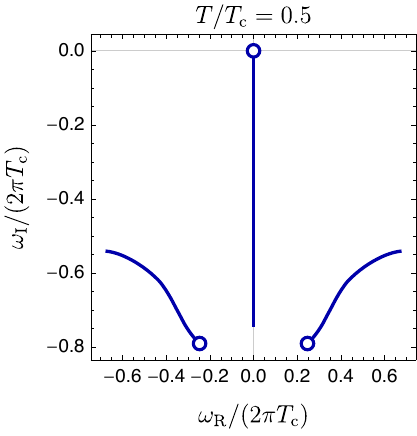}
\caption{
	Locations of the QNMs for $Z_{\chi}$ in the complex $\omega$ plane.
	The small open circles denotes the locations at $k_{x}=0$, and the curves are trajectories when $k_{x}$ increases. 
	Left: $T/T_{\rm c}=0.8$. Right: $T/T_{\rm c} = 0.5$.
}
\label{fig:qnms}
\end{figure}

Figure \ref{fig:NG_modes} shows the dispersion relations for various temperatures below and above $T=T_{\rm c}$.
The dispersion relations shown in Fig.~\ref{fig:NG_modes} are well approximated by
\begin{equation}\label{eq:dispersion_relation}
	0 = M^2 - \omega^2 - i \Gamma \omega  + \Gamma D k^2,
\end{equation}
where $M, \Gamma, D$ are real fitting parameters given by table \ref{tab:disp}.
At $T=0.8 T_{\rm c}$, in the broke phase, the mass of the NG mode $M$ vanishes.
In this case, there are two modes obeying
\begin{align}
	\omega =& -i D k_{x}^2 + \order{k_{x}^3},\\
	\omega =& -i \Gamma  +i D k_{x}^2 + \order{k_{x}^3}
\end{align}
in the long wavelength limit.
The former is a zero mode at $k_{x} = 0$ and considered as the NG mode associated with the broken $U(1)$ symmetry.
We also have the second gapped mode, which has a purely imaginary frequency at $k_{x}=0$.
Figure \ref{fig:second_NGmode} shows the behavior of the quasinormal frequency of the second mode under varying temperature.
At low temperatures, we observe merging of the second and the third modes.
It implies that the third mode is also involved in the behavior of lower QNMs when the temperature is small enough.
We obtain the temperature, at which the behavior of the QNMs is changed, as $T =0.77718 T_{\rm c}$.
Figure \ref{fig:qnms} shows the behaviors of QNMs at $T = 0.8 T_{\rm c}$ and $0.5 T_{\rm c}$.
The zero and the first imaginary modes form a pair at $T=0.8 T_{\rm c}$, whereas the two modes having complex quasinormal frequencies at $k_{x}=0$ form a pair at $T=0.5 T_{\rm c}$.
As a result, the dispersion relations cannot be described by Eq.~(\ref{eq:dispersion_relation}) for $T < 0.77718 T_{\rm c}$.
It implies that we can no longer ignore the contributions of higher-order terms in the dispersion relations.
In \cite{Ishigaki:2024djz}, a similar recombination of QNMs is observed for the vector perturbations, and the characteristic temperature was estimated as $T =0.77890 T_{\rm c}$.
We expect there is a common geometrical origin behind those QNM recombination phenomena.

\begin{table}
\centering
\begin{tabular}{c|ccc}
 $T/T_{\rm c}$ & $M/m$ & $\Gamma/m$ & $D m$\\\hline
 0.8 & 0 & 1.0356 & 0.99417 \\
 1.2 & 0.32902 & 1.0930 & 0.9476 \\
 2.0 & 0.85323 & 1.4770 & 0.73350 \\
\end{tabular}
\caption{
	Fitting parameters in Eq.~(\ref{eq:dispersion_relation}).
}
\label{tab:disp}
\end{table}

\section{Conclusion and discussion}\label{sec:discussion}
In this paper, we studied the $U(1)$ symmetry breaking in the model proposed by \cite{Ge:2023yom} under the neutral limit.
The model involves the axio-dilaton field, extended from the original 4d Gubser-Rocha model.
\rev{%
By considering the appropriate boundary action corresponding to the triple-trace deformation, the model exhibits the phase transition in the neutral limit.
}%
One of the most attractive features of this model is that the broken-phase solution is also given analytically.

The critical exponents of the global $U(1)$ phase transition in the bulk are obtained as Eq.~(\ref{eq:critical_exponents}).
The numerical result of the susceptibility is also consistent with the value of $\gamma$ under the scaling relation with $\alpha$ and $\beta$.
In contrast with the phase transition in holographic superfluid models, the exponent $\beta$ becomes $1$ rather $1/2$ in our case.
The value of $\beta = 1/2$ is a typical value of the mean-field theory, in which several critical exponents are given by
\begin{equation}
	\text{(MFT)}\quad
	\alpha = 0,\quad
	\beta = \frac{1}{2},\quad
	\gamma = 1,\quad
	\delta = 3.
\end{equation}
The upper critical dimension is four.
In holography there are a few examples where the critical exponents deviate from the mean field values.
For instance, the authors of \cite{Cai:1998ji} have shown that a rotating D3 black brane, which can be regarded as a single charge STU black hole after taking dimensional reduction, exhibits a phase transition with $\alpha=1/2$.%
\footnote{
    The dynamic critical exponent of this setup was also studied in \cite{Maeda:2008hn}.
}
The authors of \cite{Franco:2009yz} have shown that $\beta=1$ is observed in the generalized s-wave holographic superconductor model, which has cubic terms of the charged scalar in the \emph{bulk} coupling with gauge fields.
In \cite{Zeng:2010zn}, they have also obtained the result with $\beta=1$ in a similar model with a \emph{bulk} cubic potential.
\rev{
The authors of \cite{Matsumoto:2022nqu} have observed the deviation of the critical exponents from the mean-field values at a current-driven tricritical point of non-equilibrium steady states by using holography. 
}
In our case, the broken symmetry of the phase transition is clear: the global $U(1)$ symmetry of the operator dual to the axio-dilaton field in the S-type model, or $Z_{2}$ in absence of the axion.
Meanwhile, our results coincide with those of the mean-field theory of the percolation theory \cite{hara1990mfc, PhysRevE.54.6003, ellis2023fifty}.%
\footnote{\rev{
	The values of $\beta=1$ and $\gamma=1$ can be found in \cite{hara1990mfc, PhysRevE.54.6003}, which lead to $\alpha = -1$ with the scaling relation (\ref{eq:first_scaling_relation}).
	In the review \cite{ellis2023fifty}, one can also find the values of $\alpha, \beta, \gamma$.
	Note that the critical exponents involved in the hyperscaling relation, such as $\nu$, are still different.
}}
The critical exponents are given by
\begin{equation}\label{eq:exponents}
	\text{(Percolation MFT)}\quad
	\alpha = -1,\quad
	\beta = 1,\quad
	\gamma = 1,\quad
	\delta = 2.
\end{equation}
The upper critical dimension is six.
Actually, the behavior of the order parameter can be reproduced by the following cubic potential:
\begin{equation}
	V(\sigma) = V_{0} + \frac{m^2}{2} \abs{\sigma}^2 + \frac{g_{3}}{3} \abs{\sigma}^3,
\end{equation}
where $\sigma$ denotes the order parameter.
$m$ is a mass,  $g_{3}$ is a coupling constant, and $V_{0}$ is a constant energy.
By minimizing the potential, we obtain the condensate as
\begin{equation}
	\abs{\sigma} =
	\begin{cases}
		- m^2/g_{3} & m^2 < 0\\
		0 & m^2 \geq 0
	\end{cases},
\end{equation}
where we assumed $g_{3}$ is positive.
The phase transition implies $m^2 \propto (1-T/T_{c})$ in the vicinity of $T=T_{c}$. As a result, we obtain $\abs{\sigma} \propto (1-T/T_{c})$ corresponding to $\beta = 1$.
In the mean-field percolation theory, the presence of the cubic term is allowed \cite{ellis2023fifty}.
This picture might be related to that we have considered the triple-trace deformation giving an additional cubic potential term in the boundary theory.
It would be interesting to study whether it is possible to adjust the critical value by varying the way the multi-trace operator is deformed.
We leave this as a future work and an open question.

We have also checked the dynamical stability of the system in the neutral limit by analyzing QNMs associated with the axio-dilaton perturbation.
Taking into account the gauge symmetry of the system, we obtain the consistent \rev{result} with the thermodynamic analysis in \cite{Ishigaki:2024djz}.%
\footnote{
	Recently, the authors of \cite{Arean:2024pzo} investigated the (in)stability of the black hole interior in some models including the Gubser-Rocha model with explicit translational symmetry breaking.
	This (in)stability is independent of the dynamical stability governed by the spacetime outside the horizon.
}
Note that the treatment of the gauge symmetry is crucial in studying such a system coupled with the gravitational perturbations.
Incorrect treatment of the gauge symmetry can easily lead to unphysical consequences.

The dispersion relation of the NG modes is described by Eq.~(\ref{eq:dispersion_relation}), when the temperature is near to $T_{\rm c}$.
This is a typical $k$-gap behavior, which is commonly observed in a dissipative system.
According to \cite{Minami:2015uzo}, our NG mode is categorized into type A, whose dispersion relation shows the diffusive behavior.
A similar behavior was observed for chiral symmetry breaking in the D3-D7 probe brane model, where the bulk global $U(1)$ symmetry is spontaneously broken \cite{Ishigaki:2020vtr}.
In the holographic superconductors, where $U(1)$ symmetry is gauged in the bulk, the NG modes usually become second sound modes, whose dispersion relation is given by $\omega = \pm v_{\rm s} k - i D_{\rm s} k^2 + \order{k^3}$ \cite{Amado:2009ts}.
On the other hand, our results show that there is one mode with $\omega= -i D k^2 + \order{k^3}$ and another with $\omega= -i\Gamma -i D k^2 + \order{k^3}$.
These different behaviors of the NG modes may reflect whether the bulk symmetry is local or global, namely the conserved current exists or not.
The authors of \cite{Donos:2019txg} also obtained a similar diffusive dispersion relation for the broken global symmetry in the bulk.
In \cite{Amado:2013xya}, however, they obtained propagating NG modes even if the broken symmetry is ungauged but a part of the gauged non-Abelian symmetry of the holographic model.
This is probably because the global symmetry is the part of the non-Abelian symmetry.

The Gubser-Rocha model is related to eleven-dimensional supergravity theory through dimensional reduction on $S^7$.
Although the total action of the S-type model cannot be obtained from the supergravity model, the scalar potential can be derived.
As it was mentioned in \cite{Chagnet:2022yhs}, the dual field theory of the Gubser-Rocha model is probably ABJM theory \cite{Aharony:2008ug}, but the presence of the triple-trace deformation makes it difficult to understand the exact field-operator correspondence.
On the the other hand, one needs to consider another way of the dimensional reduction to obtain charged fields from the eleven-dimensional supergravity.
For instance, the authors of \cite{Gauntlett:2009zw} considered a reduction on seven-dimensional Sasaki-Einstein manifold and obtained charged scalar in the four-dimensional theory.
To study superconducting behavior from the top-down holography, the presence of the bulk charged field is essential.

Lastly, it is worth mentioning that the presence of the bulk global symmetry might be forbidden by the swampland conjecture from the viewpoint of the string theory.
However, we do not face any inconsistency at the level of the classical computation within our model.

\section*{Acknowledgments}
The authors thank Xian-Hui Ge, Zhaojie Xu, and Masataka Matsumoto for useful discussions.
This work is supported by NSFC, China (Grant No.~W2433015).

\appendix

\section{\texorpdfstring{\(\mathcal{N}=2\), \(U(1)^4\)}{N=4, U(1)\textasciicircum 4} gauged supergravity}\label{appendix:sugra}
The authors of \cite{Cvetic:1999xp, Chong:2004na} have considered $\mathcal{N}=2$, $U(1)^4$ gauged supergravity theory obtained from the maximal $\mathcal{N}=8$, $SO(8)$ gauged supergravity theory.
It is considered that the model is also obtained from the eleven-dimensional supergravity by taking dimensional reduction on $S^{7}$.
The truncated model involves three dilaton fields $\varphi_i$, axion fields $\chi_i$, and four gauge fields $A^{I}_{\mu}$.
While the total Lagrangian density is complicated, the scalar part is given by
\begin{equation}
	\frac{\mathcal{L}_{\rm scalar}}{\sqrt{-g}}
	=
	\sum_{i=1}^{3}
	\left[
		- \frac{1}{2}(\partial \varphi_{i})^2
		- \frac{1}{2}e^{2\varphi_{i}} (\partial \chi_{i})^2
		+ \frac{2}{L^2}\left(
			\cosh\varphi_{i}
			+ \frac{1}{2} \chi_{i}^2 e^{\varphi_{i}}
		\right)
	\right]
\end{equation}
If we can set $\varphi_{1} = \varphi_{2} = \varphi_{3} = \phi$ and $\chi_{1} = \chi_{2} = \chi_{3} = \chi$, it coincides with Eq.~(\ref{eq:S-completed}) except the terms of the gauge fields.
The total Lagrangian density incorporating the gauge fields can be found in \cite{Cvetic:1999xp} Eq.~(B.7).
As far as we checked, however, the above ansatz does not work for the finite three-equal charge case: $A^{1}=0$ and $A^{2}=A^{3}=A^{4}=A/\sqrt{3}$.
For example, the equations of motion for each dilaton exhibit inconsistency, such as $\pdv{\mathcal{L}}{\varphi_{1}} \neq \pdv{\mathcal{L}}{\varphi_{2}}$, under this ansatz.
Furthermore, even if we substitute the above ansatz into the action of the $U(1)^4$ theory, the gauge field terms are different from Eq.~(\ref{eq:S-completed}).
The authors of \cite{Chow:2013gba} actually obtained a general dyonic black hole solution in the $U(1)^4$ gauged supergravity.
It would be interesting to investigate the properties of their solution for the case of three equal charges, but we leave this for future work.

When $\chi=0$, the ansatz is consistent.
In this case, the Lagrangian density is reduced to that of the original Gubser-Rocha model:
\begin{equation}
	\frac{\mathcal{L}}{\sqrt{-g}} = 
	R - \frac{1}{4} e^{-\phi} F^2
	- \frac{3}{2}(\partial \phi)^2
	+ \frac{6}{L^2} \cosh\phi.
\end{equation}

\section{Changing variable for the axio-dilaton field}\label{appendix:another_action}
\begin{figure}[htbp]
	\centering
	\includegraphics[width=0.8\linewidth]{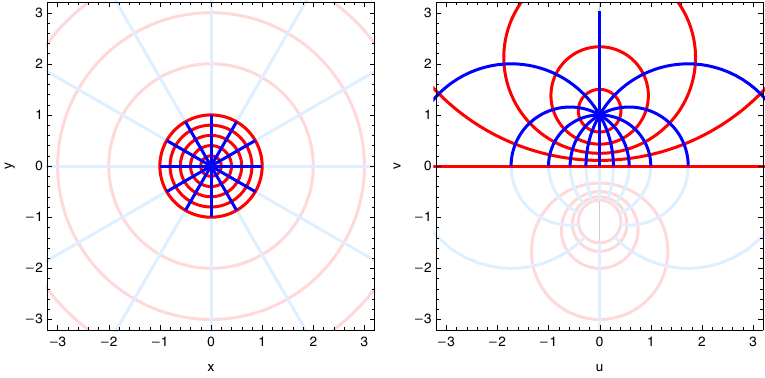}
	\caption{
		Illustration of the Cayley map between $\Phi = x+i y$ and $\tau = u+iv$ here.
		The region $|x + i y|<1$ in the left panel is mapped to the upper half plane in the right panel.
	}
	\label{fig:map_xy2uv}
\end{figure}

In this section, we show that the global $U(1)$ symmetry in our model can be regarded as rotation of the phase of the complex scalar field after changing variable of the axio-dilaton field.
The transformation is given by a specific M\"obius transformation known as the Cayley transformation:
\begin{equation}
	\tau =
	i \frac{1+i \Phi}{1-i\Phi}.
\end{equation}
One can find a similar argument in Section 15.4.1 of textbook \cite[pp.451-454]{ortin2004gravity}.
The complex map is illustrated in Fig.~\ref{fig:map_xy2uv}.
Writing the real and imaginary part as $\Phi = \Phi_{1} + i \Phi_{2}$, we obtain
\begin{equation}
	\tau_{1} = - \frac{2 \Phi_{1}}{\Phi_{1}^2 + (1+\Phi_{2})^2}
	= \frac{-2 \Phi_{1}}{|i + \Phi|^2},\quad
	\tau_{2} = - \frac{- 1 + \Phi_{1}^2 + \Phi_{2}^2}{\Phi_{1}^2 + (1+\Phi_{2})^2}
	= \frac{1 - |\Phi|^2}{|i + \Phi|^2}.
\end{equation}
The kinetic term is written as
\begin{equation}
	\frac{4}{(\tau-\bar{\tau})^2}
	\partial_{\mu} \tau \partial^{\mu} \bar{\tau}
	=
	\frac{4}{(1 - |\Phi|^2)^2}
	\partial_{\mu}\Phi \partial^{\mu}\bar{\Phi}.
\end{equation}
The Lagrangian density (\ref{eq:S-completed_tau}) is written as
\begin{equation}
	\frac{1}{\sqrt{-g}}\mathcal{L}_{1}
	=
	R
	- 6 \frac{|\partial \Phi|^2}{(1 - |\Phi|^2)^2}
	- \frac{6}{L^2} \frac{1 + |\Phi|^2}{1 - |\Phi|^2}
	- \frac{1}{4} \frac{1 - |\Phi|^2}{|i + \Phi|^2} F^2
	- \frac{1}{4} \frac{2 \Phi_{1}}{|i + \Phi|^2} F \tilde{F}.
\end{equation}
In this form, the $U(1)$ symmetry is manifestly a rotational symmetry of the phase of $\Phi$.
The S-dual transformation for $\Phi$ is expressed as $\Phi \to - \Phi$.
Thus, the S-dual symmetry is preserved if the potential term is given by an even function of $\Phi$.

\bibliography{main}
\bibliographystyle{ytphys}
\end{document}